\documentclass[11pt]{article}
\usepackage{moriond,epsfig}

\def\Journal#1#2#3#4{{#1} {\bf #2}, #3 (#4)}

\def\NIMA{{\em Nucl. Instrum. Methods} A}
\def\NPB{{\em Nucl. Phys.} B}
\def\PLB{{\em Phys. Lett.}  B}
\def\PRL{\em Phys. Rev. Lett.}
\def\PRD{{\em Phys. Rev.} D}

\def\CPC{{\em Comp. Phys. Comm.}}
\def\JHEP{{\em JHEP}}

\def\be{\begin{equation}}
\def\ee{\end{equation}}
\def\bea{\begin{eqnarray}}
\def\eea{\end{eqnarray}}

\begin{document}
\hspace*{12cm} lycen 2005-11
\vspace*{4cm}
\title{Production and decay of gluino pairs at hadron colliders \footnote{Proceedings of the Conference: ``QCD and High Energy Hadronic Interactions'', XXXXth Rencontres de Moriond, La Thuile, Italy - March 12-19, 2005}}

\author{F. Mahmoudi \& A. Deandrea}

\address{Institut de Physique Nucl\'eaire de Lyon (IPNL), Theory Group,\\
Universit\'e Claude Bernard Lyon-1,\\
4, rue Enrico Fermi, 69622 Villeurbanne cedex, France }

\maketitle\abstracts{
At hadron colliders, one of the most important channels for sparticle production is expected to be the gluino pair production. In the scenario where the sbottom is lighter than the gluino, gluinos can decay into sbottom and bottom quarks. Sbottoms can subsequently decay into bottom quarks and neutralinos. Hence, one expects a rich signature consisting of four b-jets and the missing transverse energy from neutralinos. To compute the cross section of this reaction, it is important to find tools and techniques suitable for the calculation of diagrams with many particles in the final state. We present here a way to deal with such complicated reactions.
}

\section{Introduction}
%
In the Minimal Supersymmetric extension of the Standard Model \cite{mssm1,mssm2} (MSSM), every fundamental particle must have a supersymmetric partner with spin differing by 1/2 unit. When considering the conservation of R-parity, the lightest supersymmetric particle (LSP) must be stable, and supersymmetric particles are produced only in even number. If the LSP is also neutral and colorless, its signature in detectors will only be a large missing transverse energy. One often assumes the lightest neutralino to be the LSP.\\
The main aim of present and future hadron colliders is the search for beyond Standard Model physics, and in particular Supersymmetry.
Because of the high center of mass energies (TeV scale), the final states will be very complex, and multi-particle final state reactions will be produced at a huge rate. Moreover, in order to have a good description of the physics at these scales, it is necessary to have precise theoretical predictions.\\
In this work, we consider the production and decay of gluino pairs at hadron colliders, which appears as one of the most important channels for sparticle production. We investigate ways to achieve high precision calculations for multi-particle final state amplitudes in the MSSM. In particular, while considering the complete gluino pair production and decay chain, one has a eight body amplitude which requires very complicated calculations. This amplitude is an interesting signature at Tevatron. Furthermore, it is a good example to develop a computation technology able to treat complex reactions.\\
In section 2 we give an overview of the gluino pair production. Section 3 describes the gluino decay modes and presents the topologies. The way we performed the calculation of the corresponding amplitudes is presented in section 4. Finally, conclusion and perspectives are given in section 5.
%
\section{Gluino pair production}
%
Gluinos and squarks are presently actively searched at Tevatron. Lower bounds have already been set up for the masses of gluinos and squarks \cite{abbott1,abbott2,abachi}. In particular, if gluinos and squarks have the same mass, the lower mass limit is 310 GeV. In the opposite case, if the gluino (squark) is heavier, the lower squark (gluino) mass limit is then 240 GeV.\\
The main modes for gluino pair production at hadron colliders are, in leading order of the perturbative expansion:\\
\\
$\bullet$ $q_i + \bar{q}_i \longrightarrow \tilde{g} + \tilde{g}$
\begin{figure}[!ht]
\begin{center}
\epsfig{file=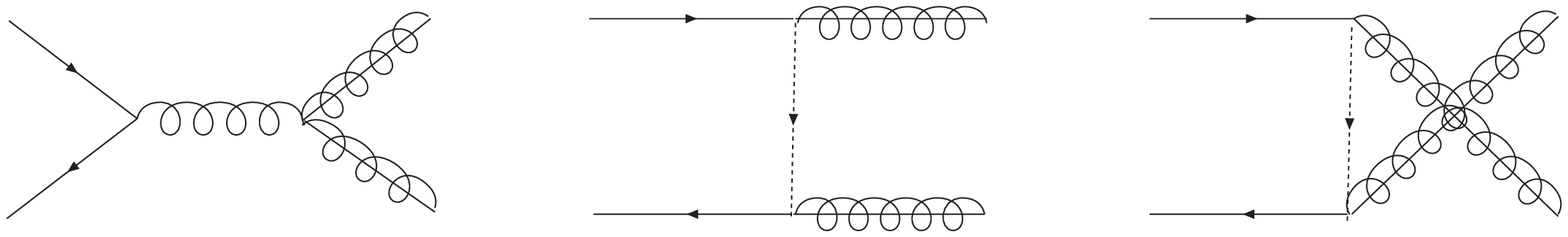,width=10cm}
\end{center}
\end{figure}\\
$\bullet$ $g + g \longrightarrow \tilde{g} + \tilde{g}$
\begin{figure}[!ht]
\begin{center}
\epsfig{file=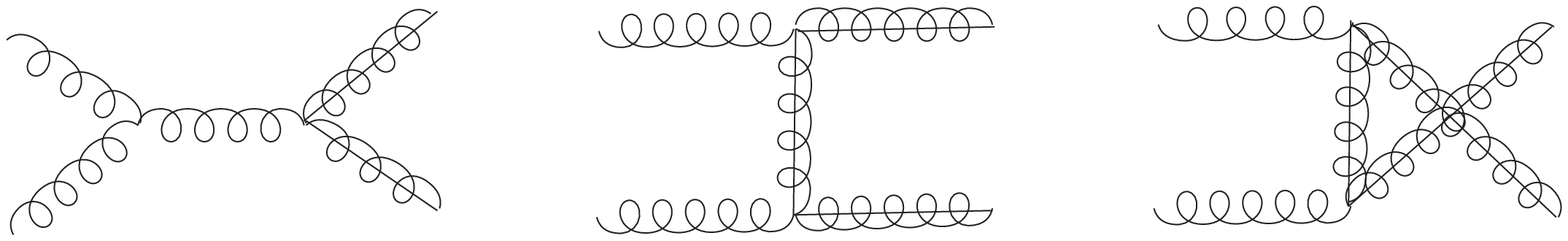,width=10cm}
\end{center}
\end{figure}\\
The relative weights of each reaction is strongly correlated to the masses of squarks and gluinos~\cite{beenakker}. Due to the presence of an extra intermediate sparticle in the t-channel, for both cases the s-channel is privileged.
%
\section{Gluino decay}
\begin{figure}
\epsfig{file=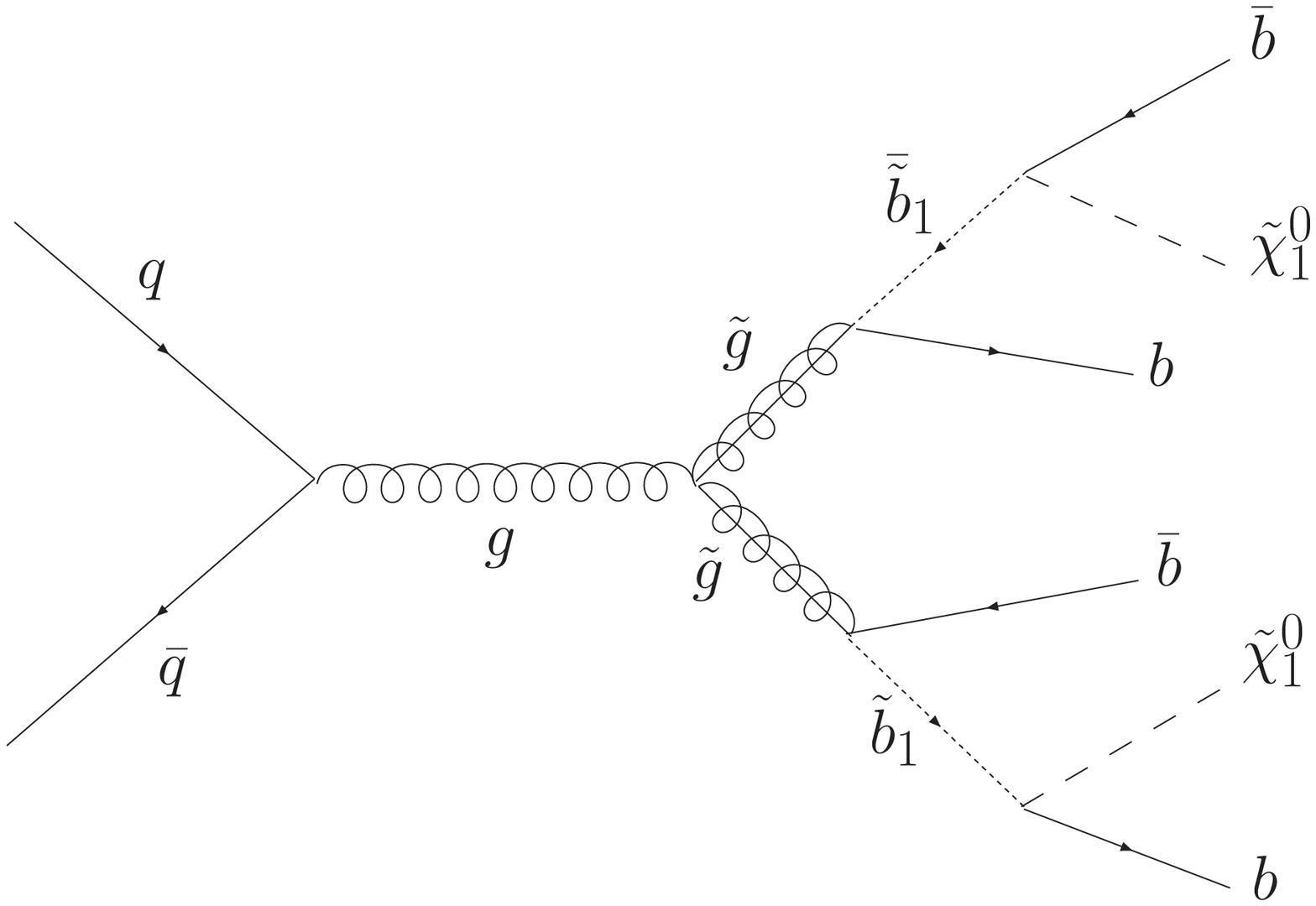,width=4cm,clip}\hspace{0.7cm}\epsfig{file=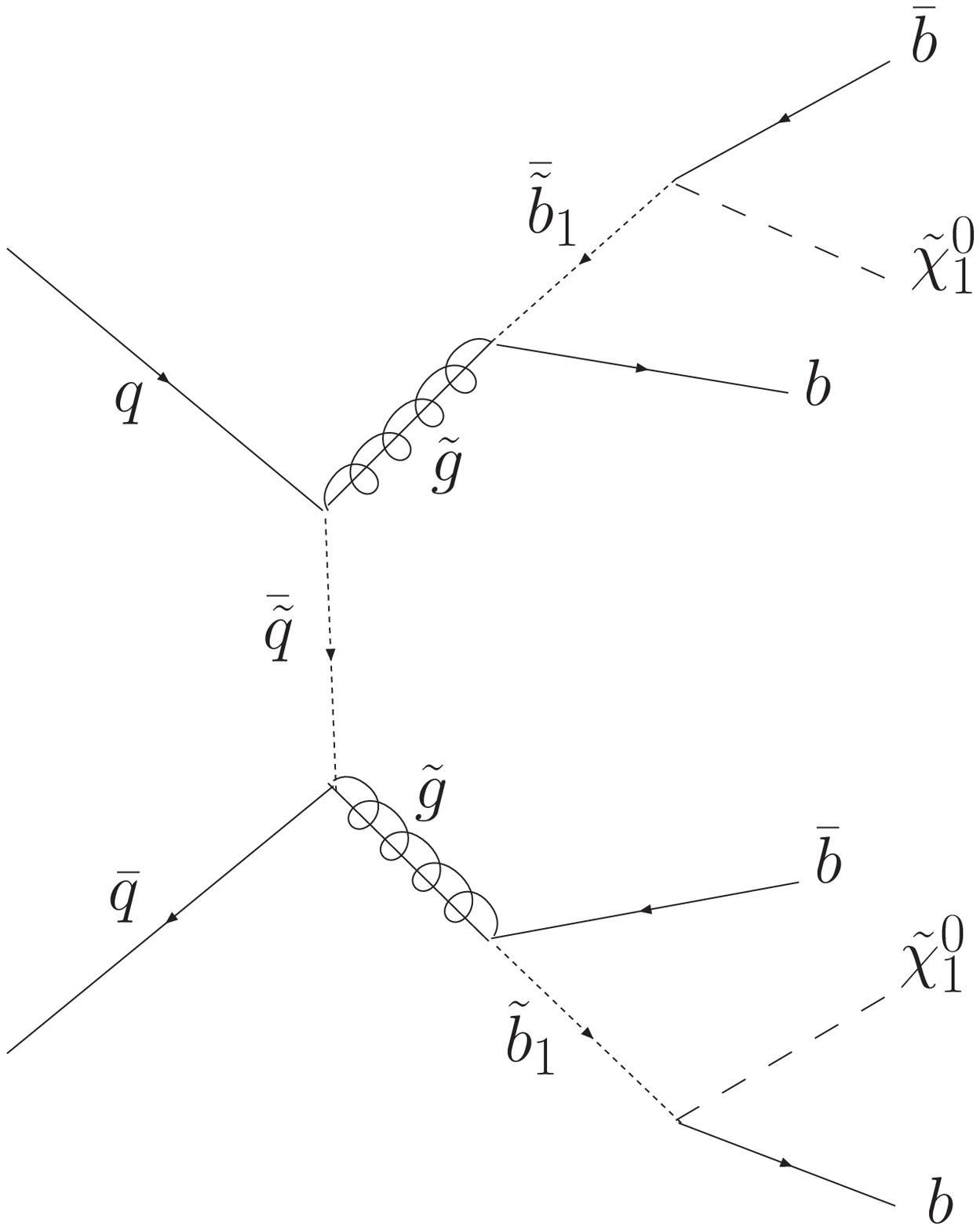,height=3cm,clip}
\hspace{0.7cm}\epsfig{file=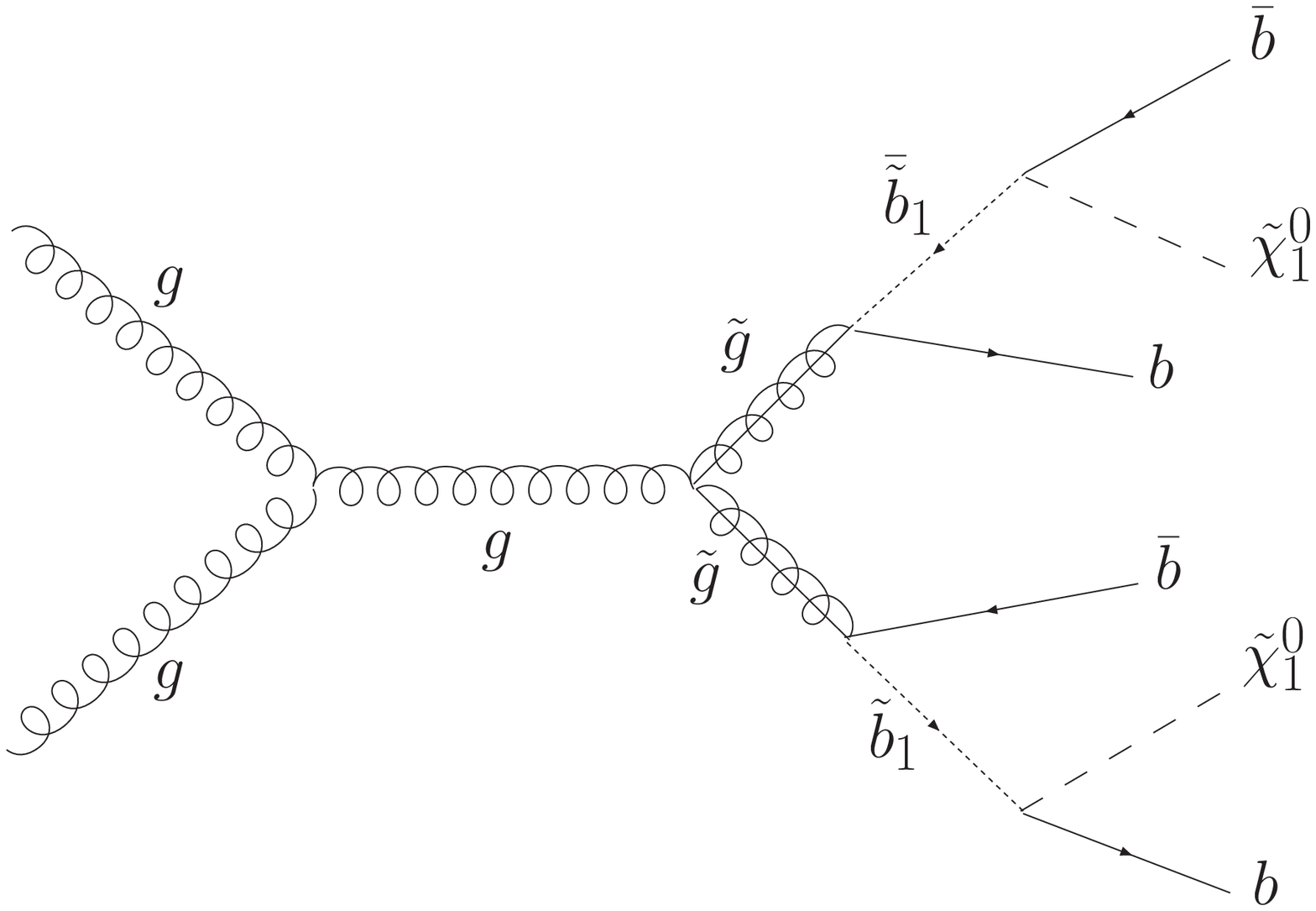,width=4cm,clip}\hspace{0.7cm}\epsfig{file=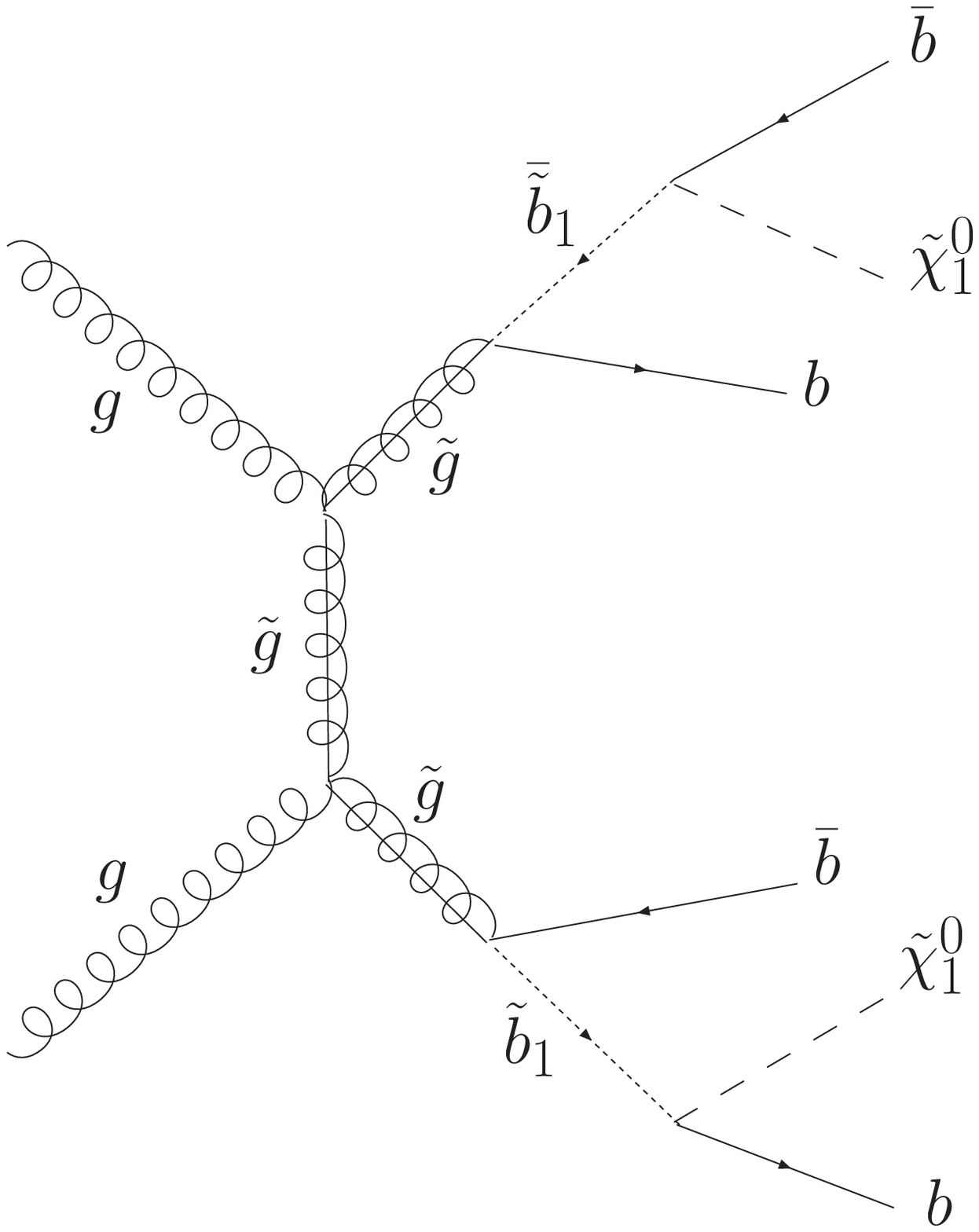,height=3cm,clip}\\
\begin{center}
(a)\hspace{3.5cm}(b)\hspace{3.3cm}(c)\hspace{3.5cm}(d)~~~~
\end{center}
\caption{Schematic representation of the topologies corresponding to $ q \, \bar{q} \, / \, g \, g \longrightarrow b \,\, \tilde{\chi}^0_1 \,\, \bar{b} \,\, b \,\, \tilde{\chi}^0_1 \,\, \bar{b}$. (a) and (b) correspond to the gluino pair production from $q \bar{q}$ annihilation, for respectively s- and t-channels. (c) and (d) correspond to the gluino pair production from $gg$, for respectively s- and t-channels.~~~~~~~~~~~~~~~~~~~~~~~~~~~~~~~~~~~~~~~~~~~~~~~~}
\label{schema}
\end{figure}

%
%
In the MSSM, for the third generation of quarks, a strong mixing can appear for the superpartner masses, which depends on the parameters of the theory. In particular, the masses of the two sbottoms read: 
\begin{eqnarray*}
m^2_{\tilde{b}_{1,2}} = \frac{1}{2} \left[ m^2_{\tilde{b}_L} + m^2_{\tilde{b}_R} 
			 \mp \sqrt{(m^2_{\tilde{b}_L} - m^2_{\tilde{b}_R})^2 + 4 m^2_b (A_b - \mu \; \tan \beta)^2 }\, \right] \;\;,
\end{eqnarray*}
where the dependence on $\tan \beta$ (ratio of the VEV's of the two Higgs fields) and $A_b$ (sbottom-Higgs trilinear coupling) appears explicitly. Thus, for a large value of $\tan \beta$, one observes a high mass splitting for the two sbottoms. This can lead to a rather low mass for the lighter sbottom, so that the $\tilde{g} \longrightarrow  b + \tilde{b}$ decay would be possible. The sbottom can then decay through  $\tilde{b} \longrightarrow  b + \tilde{\chi}^0_1$.\\
The complete reaction to be considered is therefore $ q \, \bar{q} \, / \, g \, g \longrightarrow b \,\, \tilde{\chi}^0_1 \,\, \bar{b} \,\, b \,\, \tilde{\chi}^0_1 \,\, \bar{b}$. We select the topologies represented schematically in figure~\ref{schema}, which can be experimentally identified through kinematic cuts. The experimental signature will then be the missing transverse energy of two undetected neutralinos and the presence of four b-jets.
To compute this amplitude, one has to consider all the leg permutations. This leads to 16 diagrams for each of the s-channels (a, c) and 32 diagrams for the t-channels (b, d), so that 48 diagrams have to be computed for each partonic subprocess. 
%
\section{Cross section calculations}
%
In usual methods for computing the amplitude of diagrams with many particles in the final state, one generally splits the processes into several different subprocesses, and obtains the total cross section while considering the branching ratio of each subprocess. The computations are then relatively easy, but one loses the information about particle spins and one obtains approximate results.\\
\\
On the contrary, the direct calculation is an exact method, which consists in calculating the total amplitude of a process and integrating it numerically over the phase space associated to the final state particles. This method is unfortunately difficult to follow, especially when the number of diagrams corresponding to the studied process is large or/and if the number of particles in the final state is high. To achieve those computations, several automatic codes (such as CompHEP \cite{comphep}, GRACE \cite{grace}, MadGraph \cite{madgraph}, Alpgen \cite{alpgen}, FormCalc \cite{formcalc}, ...) exist, but only a few are able to deal with multiparticle processes in the MSSM. In particular, CompHEP is almost able to treat completely $2 \longrightarrow 6$ MSSM amplitudes. Unfortunately, in this case, the number of generated terms in the output files is so high that the results cannot be used, neither analytically nor numerically.\\
We computed this amplitude with Form \cite{form} and Mathematica \cite{math}, using a modified version of the FormCalc package.\\
\\
In its standard version, FormCalc can deal with $2 \longrightarrow 3$ amplitudes within the MSSM. Practically, using first the FeynArts \cite{feynarts} part of the package one can generate Feynman diagrams and write the corresponding Feynman amplitudes. Up to this step, it is possible to generate amplitudes with as many particles in the final state as wanted, but subroutines to go further do not exist for more than three particles. In the next (internal) steps, FormCalc uses Form to perform analytical calculations of the amplitudes, and then generates a Fortran code containing the squared matrix elements. Finally, it computes the cross-sections of $1 \longrightarrow 2$, $2 \longrightarrow 2$, and $2 \longrightarrow 3$ processes.\\
\\
In order to compute our $2 \longrightarrow 6$ amplitude, we had to modify the internal files of FormCalc which are written in Form. Those modifications are suitable for any $2 \longrightarrow 6$ amplitude, and can be easily extended to even more complicated cases. The generated Fortran code for the squared matrix elements is then relatively concise and fast thanks to the use of the Weyl-van-der-Waerden formalism \cite{wvdw}. Instead of elaborating new kinematics modules for FormCalc enabling to calculate numerically the cross section, we chose rather to insert our results into PYTHIA \cite{pythia}.\\
Complete results will be presented in a future publication.

\section{Conclusion and perspectives}
%
The gluino pair production is known to be an important channel for sparticle production at hadron colliders. In this note, we considered the gluino pair production from $q \, \bar{q}$ and $g \, g$ initial states, and their decay into sbottom and bottom quarks in the MSSM. The whole reaction is then $q \, \bar{q} \, / \, g \, g \longrightarrow b \,\, \tilde{\chi}^0_1 \,\, \bar{b} \,\, b \,\, \tilde{\chi}^0_1 \,\, \bar{b}$. The direct calculation of this amplitude has been performed while modifying the FormCalc package for Mathematica. This study allowed us to get adequate tools and techniques for the direct calculation of $2 \longrightarrow 6$ amplitudes.\\
\\
The next step, in collaboration with the D0 group at IPNL, consists in a comparison of the theoretical results and the experimental data, in order to derive constraints on the gluino and sbottom masses. Moreover, it will be interesting to compare the results of the direct calculation to those of the usual method used in PYTHIA (branching ratio method) and to check whether this simplest method is reliable for complicated cases with many particles in the final state and asymmetric events.\\
\\
The technology employed for these calculations can hopefully be used for other amplitudes or for even more complicated cases. Thus, it will be very useful and interesting to use these tools to make precise predictions for Tevatron and LHC.\\
\section*{Acknowledgments}
%
I would like to thank Steve Muanza and Thomas Millet for useful discussions. I am also grateful to the organizers for this very interesting conference. Feynman diagrams in this paper are drawn with Jaxodraw \cite{jaxo}.\\

\section*{References}
%


\begin{thebibliography}{99}

\bibitem{mssm1} J. Wess and B. Zumino, \Journal{\NPB}{70}{39}{1974}.

\bibitem{mssm2} P. Fayet, \Journal{\PLB}{64}{159}{1976},\\
		P. Fayet, \Journal{\PLB}{69}{489}{1977}.

\bibitem{abbott1} B. Abbott et al., \Journal{\PRL}{82}{29}{1999}.

\bibitem{abbott2} B. Abbott et al., \Journal{\PRL}{83}{4937}{1999}.

\bibitem{abachi} S. Abachi et al., \Journal{\PRL}{75}{618}{1995}.

\bibitem{beenakker} W. Beenakker, R. Hopker, M. Spira and P.M. Zerwas, \Journal{\NPB}{492}{51}{1997}. 

\bibitem{comphep} E. Boos et al., \Journal{\NIMA}{534}{250}{2004}.

\bibitem{grace} J. Fujimoto et al., \Journal{\CPC}{153}{106}{2003}.

\bibitem{madgraph} F. Maltoni and T. Stelzer, \Journal{\JHEP}{0302}{027}{2003}.

\bibitem{alpgen} M.L. Mangano, M. Moretti, F. Piccinini, R. Pittau and A. Polosa, \Journal{\JHEP}{0307}{001}{2003}.

\bibitem{formcalc} T. Hahn, Talk given at the 7th DESY Workshop On Elementary Particle Theory, MPP-2004-71, hep-ph/0406288.

\bibitem{form} J.A.M. Vermaseren, math-ph/0010025.

\bibitem{math} S. Wolfram, ISBN 1579550223.

\bibitem{feynarts} T. Hahn, \Journal{\CPC}{140}{418}{2001}.

\bibitem{wvdw} S. Dittmaier, \Journal{\PRD}{59}{016007}{1999}.

\bibitem{pythia} T. Sj\"ostrand, P. Ed\'en, C. Friberg, L. L\"onnblad, G. Miu, S. Mrenna and E. Norrbin, \Journal{\CPC}{135}{238}{2001}.

\bibitem{jaxo} D. Binosi and L. Theussl, \Journal{\CPC}{161}{76}{2004}.

\end{thebibliography}
\end{document}